\documentclass[11pt,twoside]{article}
\usepackage{./asp2014}

\aspSuppressVolSlug
\resetcounters

\begin{document}

\title{MOSAIC: a Multi-Object Spectrograph for the E-ELT}
\author{Andreas Kelz$^1$, Francois Hammer$^2$, Pascal Jagourel$^2$ for the MOSAIC consortium
\affil{$^1$Leibniz-Institut f\"ur Astrophysik Potsdam (AIP), An der Sternwarte 16, D-14482 Potsdam, Germany; \email{akelz@aip.de}}
\affil{$^2$GEPI, Observatoire de Paris, CNRS, Univ. Paris Diderot, Place Jules Janssen, 92190 Meudon, France; \email{francois.hammer@obspm.fr}} 
}

\paperauthor{Andreas Kelz}{akelz@aip.de}{ }{Leibniz-Institut f\"ur Astrophysik Potsdam (AIP)}{3D \& Multi-Object Spectroscopy}{Potsdam}{Brandenburg}{14482}{Germany} 

\begin{abstract}
The instrumentation plan for the European-Extremely Large Telescope foresees a 
Multi-Object Spectrograph (E-ELT MOS). 
The MOSAIC project is proposed by a European-Brazilian consortium, to provide a unique MOS facility for astrophysics, studies of the inter-galactic medium and for cosmology. 
The science cases range from spectroscopy of the most distant galaxies, mass assembly and evolution of galaxies, via resolved stellar populations and galactic archaeology, to planet formation studies. 
A further strong driver are spectroscopic follow-up observations of targets  
that will be discovered with the James Webb Space Telescope. 
\end{abstract}

\section{Motivation}
The workhorse instruments of the current 8-10m class observatories are multi-object spectrographs (MOS), providing comprehensive follow-up of ground-based and space-borne imaging data. 
With the advent of even deeper imaging surveys from, e.g., HST, VISTA, JWST and Euclid, 
many science cases require complementary spectroscopy with high sensitivity and good spatial resolution to identify the objects and to measure their astrophysical parameters. 
The light-gathering power of the 39m E-ELT and its spatial resolution, combined with a MOS, will enable the large samples necessary to tackle some of the key scientific drivers of the E-ELT project, ranging from studies of stellar populations out to the highest-redshift galaxies. Consequently, a MOS-facility is foreseen within the E-ELT instrumentation plan \citep{ramsay_2014}. 

The MOSAIC consortium (lead by GEPI) proposes a MOS-instrument that is based on two previous concepts, EAGLE and OPTIMOS-EVE \citep{hammer_2014}. It foresees various observing modes that either provide a high multiplex, while using the E-ELT ground-layer adaptive optics system (GLAO), or a high-definition mode for fewer targets, but with better spatial resolution, enabled by multi-object adaptive optics (MOAO). The spectroscopy covers both the optical and the near-infrared.

\section{Science cases for an ELT-MOS} 

For the EAGLE and OPTIMOS-EVE Phase A studies, top-level scientific questions were identified. 
These were re-examined and a unified set of requirements was presented \citep{evans_2014} for a versatile MOS-instrument that would exploit both the excellent spatial resolution in the near-infrared envisaged for EAGLE, combined with aspects of the spectral coverage and large multiplex of EVE. 
Further consolidation of the science cases led to the ELT-MOS white paper \citep{evans_2015}.  
Table~1 lists the eight top-level science cases which are briefly described below. 

 \begin{table}[!ht]
\caption{The top-level science cases for MOSAIC}
\smallskip
\begin{center}
{\small
\begin{tabular}{ll}  
\tableline
\noalign{\smallskip}
SC1 	& 	Spectroscopy of the most distant galaxies  \\ 
SC2 	& 	Evolution of large-scale structures  \\ 
SC3 	& 	Mass assembly of galaxies through cosmic times  \\ 
SC4 	& 	AGN-galaxy co-evolution and AGN-feedback \\ 
SC5 	& 	Resolved stellar populations beyond the local group  \\ 
SC6 	& 	Galaxy archaeology  \\ 
SC7 	& 	Galactic centre \\ 
SC8 	& 	Planet formation in different environments  \\ 
\noalign{\smallskip}
\tableline 
\end{tabular}
}
\end{center}
\end{table}

\begin{itemize}
\item \emph{Most distant galaxies}: MOSAIC shall enable detailed studies of the very first galaxies. The analysis of this light will provide vital information for the study of the early epoch when the Universe was re-ionized and its gas changed from a universally neutral to ionized state. 

\item \emph{Evolution of large-scale structures}: MOSAIC shall map the 3d-structures of the gas between galaxies, which acts as a reservoir of matter from which proto-galaxies can form, or which can feed gas into existing galaxies, aiding star formation. Clusters of galaxies lie in the highest density regions in which MOSAIC will probe the dynamical mechanisms governing galaxy formation and evolution at different epochs.

\item \emph{Galaxy evolution with cosmic time}: MOSAIC shall dissect galaxies over the full lifetime of the observable Universe, gathering information on their physical and chemical properties. This will allow an understanding of the origins of present-day massive galaxies, such as the Milky Way or Andromeda. MOSAIC will also provide new insights into low-mass dwarfs and low-surface brightness galaxies, which are out of reach of current facilities, but may play a major role in shaping galaxy evolution.

\item \emph{AGN-galaxy co-evolution}: The correlated growth of super-massive black holes (thought to be at the centre of most present-day galaxies) and their host galaxies by some self-regulated feedback processes, is a key question for galaxy evolution models. This feedback is thought to be related to massive outflows, driven by active galactic nuclei (AGN) and supernova explosions. MOSAIC shall provide substantial samples of galaxies in which the physical and geometrical parameters of such outflows can be measured.

\item \emph{Galaxy archaeology and stellar populations in the Milky Way and beyond}: As stars encapsulate information about the chemical composition of the gas they formed from, they retain some of the history of their host galaxy. As the elements are produced on different time scales for different stars, they also provide a timing of events. MOSAIC will observe stellar populations in the Milky Way, in nearby galaxies, and bright stars out to tens of Mega parsecs and measure their elemental abundances and radial velocities. This yields direct estimates of the chemistry for a large volume of the local Universe. Some of the old and metal-poor stars will provide a connection to the first galaxies.

\item \emph{Exploring the centre of the Milky Way}: One of the most spectacular results of the past decade was the observed orbits of stars around Sgr A$\star$, the massive black hole at the centre of the Galaxy. Surrounding this central region are puzzling structures of gas, dust, and associated star formation, but these remain out of reach of current facilities. MOSAIC will provide insights into the physical conditions of this region.

\item \emph{Planet formation in different environments}: The number of known exo-solar planets  grows rapidly and demands studies regarding the importance of the environment - specifically stellar density and metallicity - on their formation. MOSAIC shall undertake comprehensive radial-velocity studies of stars in considerably more diverse environments than currently possible, e.g. in open and globular clusters, spanning a wide spread of densities and  metallicities, at a range of distances from the centre of the Galaxy.
\end{itemize}

\section{Technical Concept for MOSAIC} 

\articlefigure[width=0.70\textwidth]{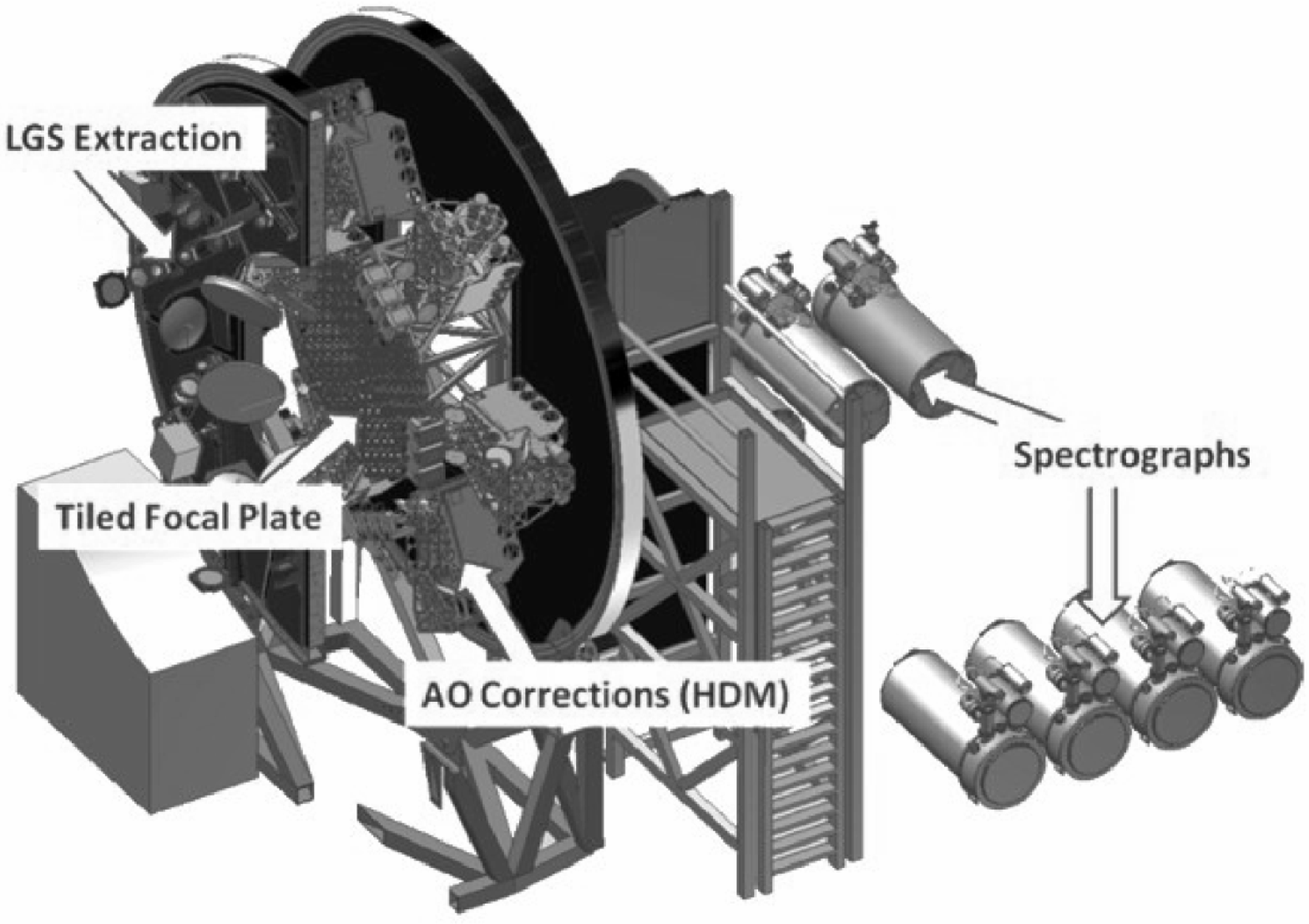}{fig1}
{Design view of the current MOSAIC concept at the Nasmyth platform of the E-ELT (the light from the telescope enters from the left). The focal plate features multi-object adaptive optics modules and is mapped by hundreds of tiles to account for the non-telecentricity of the current ELT optical design. The spectrographs are mounted to the stable platform and linked by optical fibre bundles.}  

The design of a MOS-instrument that preserves and combines the capabilities of the previous EAGLE and EVE concepts and simultaneously complies with the conditions set by the E-ELT, such as its non-telecentricity, is highly complex. In the current design concept (Fig. \ref{fig1}), MOSAIC features a focal plate with 200 tiles that carry deployable elements (fibre bundles and mirrors) to pick-up the AO-corrected light. The Adaptive Optics (MOAO) correction modules (for HDM) and the natural guide star sensing devices are at the focal plate, while the laser guide star modules are further upstream. Altogether, eight spectrographs (5 for the visible, 3 for the near-infrared) are located at the gravity-stable Nasmyth platform and fed by optical fibres. This combination enables the three main observing modes, listed in Table 2.

\begin{table}[!ht]
\caption{MOSAIC observing modes}
\smallskip
\begin{center}
{\small
\begin{tabular}{ll}  
\noalign{\smallskip}
\tableline
\noalign{\smallskip}
{\bf High-Multiplex Mode} 	& {\bf (HMM)} \\ 
Multiplex		& 	200 \\ 
Spatial aperture 	&	0.9 arcsec  \\  
Wavelength coverage 	& 	400-1800 nm  \\ 
Spectral Resolution 	& 	5000 and 15,000 \\ 
\tableline
{\bf High-Definition Mode} 	& {\bf (HDM)} \\ 
Multiplex		& 	10 IFUs \\ 
IFU field of view 	& 	2 $\times$ 2 arcsec \\ 
Spaxel size 		&	75 milli arcsec  \\  
Wavelength coverage 	& 	800-1800 nm  \\ 
Spectral Resolution 	& 	5000  \\ 
\tableline
{\bf Inter Galactic Medium}  	& {\bf (IGM)} \\ 
Multiplex		& 	10 IFUs \\ 
IFU field of view 	& 	2 $\times$ 2 arcsec  \\ 
Spaxel size 		&	0.3 arcsec  \\  
Wavelength coverage 	& 	400-1000 nm  \\ 
Spectral Resolution 	& 	5000 \\ 
\tableline
\noalign{\smallskip}
\end{tabular}
}
\end{center}
\end{table}

\subsection{Fiber-studies for the Science Optical Signal Transport System} 

MOSAIC will use miniaturized (micro-lens coupled) fibre-bundles of various sizes and multiplex. During prototype testing, the physical properties, the relative throughput and the focal ratio degradation (FRD) of such deployable bundles was studied \citep{guinouard_2012}, \citep{kelz_2014}. While the use of fibre-bundles has many advantages, such as target allocation flexibility and remote spectrograph location, their performance and stability is crucial, e.g. for accurate sky background subtraction \citep{puech_2014}.


\acknowledgements AIP gratefully acknowledges support through the funding of the BMBF Verbundforschung grant no. 05A14BA1.



\end{document}